# High-Curie-temperature ferromagnetism in bilayer $CrI_3$ on bulk semiconducting substrates

Nanshu Liu, Si Zhou[*], Jijun Zhao

*Key Laboratory of Materials Modification by Laser, Ion and Electron Beams (Dalian University of Technology), Ministry of Education, Dalian 116024, China*

## Abstract

Two-dimensional (2D) ferromagnetic (FM) semiconductors with high Curie temperature have long been pursued for electronic and spintronic applications. Here we provide a general strategy to achieve robust FM state in bilayer $CrI_3$ of the monoclinic stacking, which intrinsically has interlayer antiferromagnetic (AFM) order and weak in-plane FM coupling. We showed that the proximity effect from bulk semiconducting substrates induces electronic doping and significantly increases the FM nearest-neighbor exchange for bilayer $CrI_3$, leading to the AFM-to-FM transition for the interlayer spin configuration as well as enhanced intralayer FM coupling. By first-principles calculations and Monte Carlo simulations, bulk and 2D semiconductors providing different interaction strengths from strong covalent bonding to weak van der Waals (vdW) interaction with $CrI_3$ are compared to thoroughly address the substrate effect on magnetic behavior and Curie temperature of bilayer $CrI_3$. These theoretical results offer a facile route for direct synthesis of 2D ferromagnets on proper semiconducting substrates to achieve high Curie temperature for device implementation.

---
[*] Corresponding author. Email: sizhou@dlut.edu.cn (S. Zhou)

# 1. Introduction

Two-dimensional (2D) magnetic materials have attached tremendous attentions owing to their peculiar spin-related phenomena and immense potentials for magneto-optics, magneto-electronics and spintronic devices [1−5]. As a rising star of 2D ferromagnets, monolayer $CrI_3$ possesses many exciting features including perfect crystalline order, intrinsic Ising type ferromagnetism [6,7], strong magnetic anisotropy, and a sizable band gap of about 1.2 eV [8]. Unfortunately, the ferromagnetic (FM) coupling is rather weak in monolayer $CrI_3$ with a Curie temperature ($T_c$) of only 45 K, while bilayer $CrI_3$ with monoclinic stacking even exhibits antiferromagnetic (AFM) behavior with opposite magnetic moments from the neighboring FM layers [9]. Clearly, the fragile magnetic order of $CrI_3$ susceptible to the variation of layer thickness and stacking order has greatly hindered the application of this novel 2D material.

For practical interests, large-area $CrI_3$ sheets on semiconducting substrates are highly desirable, and an ideal scenario for device implementation is to select proper substrates to achieve robust FM order in $CrI_3$ ultrathin films. Currently, monolayer and multilayer $CrI_3$ can be synthesized by the exfoliation method and then transferred to insulating substrates, as well as be epitaxially grown on metal and nonmetal substrates, such as gold [10], highly oriented pyrolytic graphite (HOPG) [6] and yttrium stabilized zirconia [11]. Generally speaking, most 2D materials as substrates provide van der Waals (vdW) interaction with $CrI_3$ and can either retain or slightly perturb its intrinsic electronic band structure. In contrast, bulk materials with polar surface and dangling bonds may covalently couple to $CrI_3$, resulting in some exotic functionalities at the interfaces. Therefore, appropriate choice of substrates enables a large degree of freedom for band structure engineering of 2D $CrI_3$ and may give rise to unprecedented device architectures.

Previous studies have explored the proximity effect in several $CrI_3$-based vdW heterostructures. In experiment, the heterostructure of monolayer $WSe_2$ and few-layer $CrI_3$ has been exploited for valley manipulation of $WSe_2$ through control of the magnetization in $CrI_3$ [12]; in turn, the spin-valley properties render $WSe_2$ a magnetic sensor for mapping out the layered antiferromagnetic domain in the $CrI_3$ sheet [13,14].

On the theoretical aspect, the mechanism for valley splitting in WSe$_2$/CrI$_3$ heterostructures has been elucidated by considering the effects of stacking geometry [15], electrical field [16], and intrinsic point defects in CrI$_3$ [17]. Quantum anomalous Hall effect was also proposed by coupling Bi$_2$Se$_3$ or MnBi$_2$Te$_4$ with CrI$_3$ monolayer, which induces sizable spin splitting of the topological states [18,19]. Moreover, it was predicted that the in-plane FM coupling of CrI$_3$ monolayer can be enhanced with $T_c$ up to 85 K when supported on MoTe$_2$ [20], silicene and germanene monolayers [21], owing to spin-dependent charge transfer at the heterojunctions and introduction of extra spin superexchange path.

Directly growing or relocating CrI$_3$ sheet on proper substrates for magnetism engineering can overcome the limitations inherent to doping, functionalization, and applying external fields, which may result in unintended clustering, disorder, or phase transition. However, current understanding of the proximity effect in CrI$_3$ is based only on the aforementioned vdW heterostructures. How does the interlayer exchange of CrI$_3$ vary with the bonding nature and strength of substrate? Is it possible to retain the magnetic order and semiconducting character of CrI$_3$ covalently bound on a bulk substrate? Is there an optimal CrI$_3$-substrate interaction strength that leads to robust FM coupling between CrI$_3$ layers and high $T_c$ for practical uses? These intriguing questions are all awaited to be solved.

In this work, we explored bilayer CrI$_3$ on a variety of non-metal substrates with a wide range of interfacial coupling strength, including bulk semiconductors (II-VI compounds, silicon) as well as 2D monolayer materials (group III and group IV monochalcogenides, transition metal dichalcogenides, phosphorene, etc.). By first-principles calculations, the geometries, electronic band structures, and magnetic behavior of the supported CrI$_3$ sheet were systematically investigated. Strong FM order of bilayer CrI$_3$ with $T_c$ above 130 K was realized on selected bulk substrates, and the underlying exchange mechanism was elucidated. We further determined the key parameters that govern the exchange energy of bilayer CrI$_3$ on various substrates, paving a way for precisely modulating the magnetism of CrI$_3$ ultrathin sheets.

## 2. Computational methods

The structure, electronic and magnetic properties of bilayer CrI$_3$ on various substrates were investigated by density functional theory (DFT) calculations implemented in the Vienna *ab initio* Simulation Package (VASP) [22]. We adopted the planewave basis set with energy cutoff of 500 eV, projected augmented wave (PAW) potentials for electron-ion interaction [23], and generalized gradient approximation parameterized by Perdew, Burke and Ernzerhof (GGA-PBE) for the exchange and correlation functional [24]. Grimme's semiempirical DFT-D3 scheme for dispersion correction was employed to describe the van der Waals (vdW) interactions between CrI$_3$ layers and substrate [25]. Our test calculations show that the choice of vdW functionals may affect the absolute value of interlayer exchange energy ($E_M$) defined by Eq. (2), but give the same trend of $E_M$ for different heterostructure systems (Table SI of Supplemental Material, SM). Moreover, the ground state of freestanding bilayer CrI$_3$ of the high-temperature (HT) phase is correctly predicted to be antiferromagnetic by using PBE-D3, vdW-optB86b, and vdW-PBE functionals, all consistent with the previous theoretical results based on PBE-D3 [26]. A Hubbard on-site Coulomb parameter $U = 3.0$ eV was chosen for Cr atoms to account for the strong correlation effect, which can correctly predict the antiferromagnetic state for freestanding bilayer CrI$_3$ in the monoclinic stacking according to our calculations (see Fig. S1 of SM for details) as well as the previous theoretical study [27].

The supercells for bilayer CrI$_3$ on various substrates have lateral dimensions ($L$) of 6.76 ~ 21.20 Å and a vacuum space of 15 Å in the out-of-plane direction (Table SII). The details of how to construct the heterostructure models are described in Fig. S2 of Supplemental Material. Among the substrates, the II-VI semiconductors ZnO, ZnS and CdSe of wurtzite phase with the exposed (0001) surface were modeled by a four-layer slab, and the bottom-layer chalcogen atoms were terminated by H atoms. The in-plane lattice of CrI$_3$ sheet was slightly stretched or compressed to fit that of the substrate (with lattice mismatch below 1.3%) to ensure that the strain effect on the magnetic properties of CrI$_3$ sheet is negligible compared with the substrate effect. The Brillouin zones of the supercells were sampled by the Monkhorst-Pack **k**-point meshes with a separation of 0.01 Å$^{-1}$. With fixed supercells, all the model structures were fully relaxed for the ionic and electronic degrees of freedom with convergence criteria of 10$^{-5}$ eV for energy and 0.02 eV/Å for force, respectively. The charge transfer between CrI$_3$ sheet and substrate was evaluated by the Bader charge analysis [28].

## 3. Results and discussion

We considered bilayer CrI$_3$ with the monoclinic lattice and *C*2/*m* space group symmetry, which is the HT phase observed in experiment at 210–220 K. It is formed by two CrI$_3$ layers in AA stacking with one layer sliding by *a*/3 along the in-plane lattice vector ***a*** with respect to the other (Fig. 1(a), (b)). Within CrI$_3$ monolayer, each Cr atom is coordinated by six I atoms with the Cr−I−Cr bond angle close to 90°, resulting in the FM coupling according to the superexchange theorem [29−31]. The magnetic moment is 3 $\mu_B$ per Cr atom, and the band gap is 1.17 eV predicted by the PBE functional, as revealed by the electronic band structure and density of states (DOS) in Fig. 1(c). Under the octahedral crystal field, the *d* orbitals of Cr atoms split into threefold occupied $t_{2g}$ ($d_{xy}$, $d_{yz}$, and $d_{xz}$) orbitals and twofold unoccupied $e_g$ ($d_{z^2}$ and $d_{x^2-y^2}$) orbitals, corresponding to $t_{2g}$ ($d_{xy}$, $d_{x^2-y^2}$, and $d_{z^2}$) and $e_g$ ($d_{yz}$ and $d_{xz}$) orbitals in the present coordination system. By stacking two CrI$_3$ layers into the HT phase, $e_g$ or $t_{2g}$ states in one layer for a given spin channel is degenerate with the same band of opposite spin in the other layer [32], leading to antiferromagnetic order (Fig. 1(d)). Our calculations show that the AFM state of bilayer CrI$_3$ is slightly lower in energy than the FM state by −0.24 meV per CrI$_3$ formula unit (f.u.), consistent with the previous experimental observation and theoretical reports (−0.20 meV/f.u. [6,33−35]). As displayed by Fig. 5(c) (AFM1), the spin moments align ferromagnetically within each CrI$_3$ layer but point to the opposite directions between the layers. This is because the interlayer FM coupling is dominated by the virtual excitation from $t_{2g}$ to $e_g$ orbital, which encounters a large energy separation of 1.33 eV and cannot compete with the AFM exchange governed by interlayer $e_g$-$e_g$ and $t_{2g}$-$t_{2g}$ hybridization [26].

To enhance the $e_g$-$t_{2g}$ interaction and achieve robust FM order in bilayer CrI$_3$, we take advantage of the proximity effect by placing the CrI$_3$ sheet on a suitable substrate.

To explore the specific influence of substrate interaction on the magnetic state of the supported CrI$_3$ bilayer, we considered a variety of substrate materials, including (0001) surface of II-VI wurtzite compounds CdSe, ZnS and ZnO, bulk Si(111) surface, 2D group III monochalcogenide GaSe and InSe, 2D group IV monochalcogenide SnS, 2D transition metal dichalcogenide MoS$_2$ and WSe$_2$, phosphorene (Black P), and $h$-BN. The geometrical structures of these hybrid systems are displayed in Fig. S3. For clarity, we define the bottom CrI$_3$ layer close to the substrate as the 1$^{st}$ layer, and the top one as the 2$^{nd}$ layer. The interaction between the CrI$_3$ sheet and a substrate is characterized by the interfacial binding energy ($E_{bind}$) defined as:

$$E_{bind} = (E_{tot} - E_{CrI3} - E_{sub})/N_{Cr} \qquad (1)$$

where $E_{CrI3}$ and $E_{tot}$ are the energies of freestanding and supported bilayer CrI$_3$, respectively; $E_{sub}$ is the energy of the standalone substrate; $N_{Cr}$ is the number of CrI$_3$ formula units (f.u.) in monolayer CrI$_3$ within the supercell model (which equals to the number of Cr atoms at the interface with substrate). The interlayer distance $d$ is defined as the average vertical distance between the bottom I layer of CrI$_3$ and the substrate surface.

As revealed by Table I, the CrI$_3$ bilayer exhibits strong interaction with the II-VI semiconductor surfaces, having interlayer distances of 2.13 ~ 2.59 Å and $E_{bind}$ ranging from −2.32 eV/f.u. to −4.48 eV/f.u., while the geometric structure of the supported CrI$_3$ sheet remains almost the same as that of the freestanding bilayer (see the geometrical parameters in Table SII). At the interface, the I atoms of CrI$_3$ form covalent bonds with the surface metal atoms of the substrate, accompanied by charge transfer (CT) of 0.24 ~ 0.41 $e$/f.u. from substrate to CrI$_3$. Accordingly, differential charge density in Fig. 4(d) shows notable electron accumulation in the interfacial region. On the Si(111) surface, the substrate interaction is relatively weaker with $E_{bind}$ = −1.76 eV/f.u. and CT = 0.19 $e$/f.u., while vdW interaction is found between CrI$_3$ layer and all the considered 2D materials, manifested by their large interlayer distance (3.18 ~ 3.69 Å) as well as small $E_{bind}$ (−0.28 ~ −0.41 eV/f.u.) and CT (≤ 0.06 $e$/f.u.). For all the systems, the monoclinic stacking of bilayer CrI$_3$ is well retained, and the optimized structures of supported CrI$_3$ sheets do not show any reconstruction owing to the small lattice mismatch with various selected substrates. However, the exchange interaction between two CrI$_3$ layers is effectively modulated by the substrate, and consequently the preferred magnetic order of bilayer CrI$_3$ would be modified as will be discussed in the following.

The band structures and DOS of bilayer CrI$_3$ on various substrates are presented in Fig. 2 and Fig. S4, respectively. Significant changes to band structure have been induced by the four bulk semiconductor substrates, while the 2D materials with vdW interaction only shift the bands of CrI$_3$ by some extent. Taking bilayer CrI$_3$ on CdSe(0001) as an example (Fig. 2(a) and Fig. 3(d)), the spin-down channel remains insulating. For the spin-up channel, the top CrI$_3$ layer presents a band gap of 1.18 eV, while the bottom CrI$_3$ layer strongly coupled to the substrate shows prominent impurity states near the Fermi level, resembling the band structure character of *n*-doped CrI$_3$ [8]. The substrate-induced doping introduces itinerant carriers, giving rise to a half-metallic state for the supported bilayer CrI$_3$. The $e_g\uparrow$ orbital of the bottom CrI$_3$ layer gets closer to the $t_{2g}\uparrow$ orbital of the top layer. Due to the enhanced FM $e_g$-$t_{2g}$ hybridization, bilayer CrI$_3$ on the bulk semiconductors favor a FM ground state.

To characterize the magnetic coupling strength, we calculate the interlayer exchange energy as the energy difference between the interlayer AFM and FM spin configurations:

$$E_M = (E_{AFM} - E_{FM})/N_{Cr} \qquad (2)$$

where a negative value of $E_M$ indicates an AFM ground state. As given by Table I, the strongest FM order is obtained for the bilayer CrI$_3$ sheet on CdSe(0001) with $E_M = 9.42$ meV/f.u., almost 40 times of the magnitude of the AFM coupling for freestanding bilayer CrI$_3$ ($E_M = -0.24$ meV/f.u.), signifying much enhanced FM coupling due to substrate interaction. The other semiconductor surfaces like ZnS(0001), ZnO(0001) and Si(111) also endow FM spin configuration to bilayer CrI$_3$ with $E_M = 3.26 \sim 5.59$ meV/f.u. On some 2D materials including SnS, GaSe, MoS$_2$ and WSe$_2$, the vdW interaction causes remarkable shift of the bands of CrI$_3$, i.e. the Fermi level approaching the bottom of conduction band, while the band gap is slightly reduced to $0.44 \sim 0.96$ eV, leading to relatively weak FM order with $E_M = 0.06 \sim 2.00$ meV/f.u.. On the rest 2D materials like InSe, Black P and *h*-BN, the band gap of the CrI$_3$ sheet is almost intact, and the interlayer AFM coupling is maintained with $E_M = -0.06 \sim -0.30$ meV/f.u..

For the substrates providing vdW interaction, we further explored strategies to strengthen the interlayer ferromagnetic order of bilayer $CrI_3$ on the vdW substrate. We considered bilayer $CrI_3$ on graphene (lattice mismatch is 0.38%) as a representative system feasible in the experiment and obtained $E_M = 2.31$ meV/f.u., compared with $E_M = -0.24$ meV/f.u. for freestanding $CrI_3$ bilayer with AFM order. This means that an AFM-to-FM transition can be realized by placing bilayer $CrI_3$ on graphene as a vdW substrate. By applying an external electric field perpendicular to the $CrI_3$ sheet, $E_M$ further increases to 2.93 meV/f.u. at $F = 0.2$ V/Å (see Fig. S5(a)). Moreover, considering the possible in-plane strain in $CrI_3$ bilayer supported on certain substrate due to lattice mismatch, we examined the effect of biaxial strain on the magnetism of freestanding bilayer $CrI_3$. As shown in Fig. S5(b), the freestanding system remains AFM under stretching strain, while it switches to the FM state under compressive strain larger than 4%. In particular, $E_M$ reaches 0.45 meV/f.u. under 5% compressive strain. Therefore, by depositing HT-phase $CrI_3$ bilayer on a suitable vdW substrate to induce proper in-plane strain and meanwhile by applying a vertical electric field, it is possible to significantly enhance the interlayer FM order and raise the Curie temperature for device application.

The AFM to FM transition in bilayer $CrI_3$ is attributed to its electron transfer with the substrate, which alters not only the energy level but also the occupancy of $e_g$ orbital of the bottom $CrI_3$ layer, as illustrated by Fig. 3. The $e_g$-$t_{2g}$ interaction strength that dominates the FM coupling of $CrI_3$ layers can be reflected by the virtual exchange gap ($G_{ex}$) between the $e_g$ and $t_{2g}$ orbitals. Generally speaking, the stronger $CrI_3$-substrate binding associated with the larger interfacial charge transfer results in the smaller $G_{ex}$ and thus larger $E_M$ as well as stronger FM order, as displayed in Fig. 4(a). In particular, bilayer $CrI_3$ on II-VI semiconductors and Si(111) surface have $G_{ex}$ value reduced to $0.52 \sim 0.87$ eV, which is a significant impact on the electronic band structures, while 2D vdW substrates only induce band shifting and perturbation to the $e_g$-$t_{2g}$ interaction ($G_{ex} = 1.10 \sim 1.44$ eV, compared with 1.33 eV for the freestanding system). Moreover, the electrons transferred from the substrate occupy the $e_g$ orbitals of the bottom $CrI_3$ layer, and the corresponding occupancy can be calculated by:

$$N_{\text{occ.}} = \frac{\int_{-\infty}^{E_F} D^{e_g}(E)\,dE}{\int_{-\infty}^{+\infty} D^{e_g}(E)\,dE} \tag{3}$$

where $D^{e_g}(E)$ is the density of states from the $e_g$ orbitals of the bottom CrI$_3$ layer at a given energy $E$. Remarkably, for bilayer CrI$_3$ sheets on the bulk semiconductors, the considerable charge transfer leads to the occupation of $e_g$ orbitals to be $N_{\text{occ.}} = 6.03 \sim 12.60\%$ (Table SII), while the freestanding bilayer and that on 2D vdW substrates have small CT < 0.06 $e$/f.u. and thus $N_{\text{occ.}} = 0$. As shown in Fig. 4(b), $E_M$ generally increases with $N_{\text{occ.}}$, as larger occupation of the $e_g$ orbital of the bottom CrI$_3$ layer corresponds to more substrate-induced itinerant electron carriers, and thus leads to enhanced stability of interlayer FM spin configuration against AFM one.

The interfacial charge transfer between bilayer CrI$_3$ and substrate can be correlated with the work function ($\Phi$) of the substrate. As illustrated by Fig. 4(c) and (d), the substrates with lower $\Phi$ result in larger CT to the CrI$_3$ sheet, accompanied by stronger interfacial binding and smaller interlayer distance ($d$ < 2.60 Å), which in turn leads to the increase of $E_M$ for interlayer FM exchange. In particular, the (0001) polar surfaces of bulk II-VI semiconductors have a much smaller theoretical work function of 3.60 ~ 3.81 eV than that of 5.63 eV for freestanding bilayer CrI$_3$, thereby inducing prominent charge transfer through formation of the interfacial covalent bonds. In contrast, 2D materials have relatively larger work function in the range of 4.33 to 5.75 eV, and they donate to or even gain a small amount of electrons from the CrI$_3$ sheet. Therefore, they have less impact on the magnetic order in bilayer CrI$_3$. In other words, by choosing proper substrates based on the work function, it is possible to control the amount of charge transfer to bilayer CrI$_3$ and tune the $e_g$-$t_{2g}$ hybridization, ultimately modulating the interlayer spin configuration and exchange energy of the bilayer CrI$_3$ sheet.

To gain deeper insights into the mechanism of enhanced interlayer ferromagnetism of CrI$_3$, we considered a Heisenberg model with the following Hamiltonian

$$H = -\sum_{ij} J_{ij} M_i M_j \tag{4}$$

where $M_i$ and $M_j$ represent the magnetic moments at sites $i$ and $j$, respectively, and we used the computed magnetic moments for each CrI$_3$/substrate model from DFT calculations (see Table SII in the SM for the values of $M$); $J_{ij}$ is the exchange parameter between sites $i$ and $j$. Later, we will use the subscript ⊥ and ∥ to denote the interlayer and intralayer exchange parameters, respectively. As illustrated by Fig. 5(b), a total of five Cr−Cr neighbors in bilayer CrI$_3$ are considered — three for in-plane direction and two for vertical direction (the latter includes four nearest neighbors and four second nearest neighbors). The five lowest-energy AFM spin configurations are determined for freestanding and supported bilayer CrI$_3$, as displayed in Fig. 5(c) and Fig. S6. According to the exchange energies of these AFM states, $J_{ij}$ can be calculated by Eq. (4) (see Eq. (S1 ~ S12) in the SM for more details). For freestanding CrI$_3$ bilayer, we obtained the interlayer nearest-neighbor $J_{1\perp}$ = 0.88 meV representing the FM coupling, while the second nearest-neighbor $J_{2\perp}$ = −2.03 meV is obtained for AFM coupling, consistent with the previous theoretical reports [26]. Upon substrate interaction, $J_{1\perp}$ is generally increased and $J_{2\perp}$ usually becomes less negative for the considered substrates, that is, stronger FM interlayer nearest-neighbor exchange and weaker AFM second nearest-neighbor exchange, which explains the enhanced interlayer FM coupling for the supported CrI$_3$ sheet. In particular, $J_{1\perp}$ is notably enlarged for bilayer CrI$_3$ on II-VI semiconductors and Si(111) surface (6.32 ~ 9.19 meV), which is about 7 ~ 10 times larger than that of the freestanding form, while $J_{2\perp}$ is less affected (−1.17 ~ −4.00 meV). Fig. 5(a) reveals that the exchange energy $E_M$ for the interlayer FM spin configuration increases with $J_{1\perp}$ for various supported bilayer CrI$_3$ systems. Therefore, the CrI$_3$-substrate interaction mainly stabilizes the FM interlayer nearest-neighbor exchange term, endowing robust FM coupling between CrI$_3$ layers.

The substrates can not only trigger the interlayer AFM-to-FM transition in CrI$_3$, but also enhance the stability of intralayer FM state. As listed in Table I, for all the supported bilayer CrI$_3$ systems, the intralayer ground-state spin configuration is FM order. The bottom CrI$_3$ layer adjacent to the substrate has the intralayer exchange energy $E_{M\parallel}$ = 17.14 ~ 43.31 meV/f.u., compared with 16.30 meV/f.u. for the upper layer and 15.56 meV/f.u. for freestanding CrI$_3$ monolayer. Accordingly, the intralayer nearest-neighbor exchange parameter $J_{1\parallel}$ notably increases to 2.74 ~ 6.42 meV for the bottom CrI$_3$ layer on various substrates owing to the enhanced $e_g$-$t_{2g}$ hybridization ($J_{1\parallel}$ = 1.72 meV for freestanding CrI$_3$ monolayer). The second nearest-neighbor $J_{2\parallel}$ is also FM and overall has less change with respect to that of freestanding CrI$_3$ monolayer ($J_{2\parallel}$ = 1.54 meV). Considering two in-plane Cr−Cr near neighbors in Eq. (4), the exchange energy $E_{M\parallel}$ of intralayer FM order follows a linear relation with $J_{1\parallel}$ for the bottom CrI$_3$ layer on substrate, as revealed by Fig. 5(b).

Since bilayer CrI$_3$ involves two macroscopic spins [33], the energy required to flip a spin requires overcoming not only the interlayer exchange energy, but also the intralayer exchange energy. Therefore, the enhancement of both in-plane and vertical FM coupling for CrI$_3$ bilayer on semiconducting substrates would effectively elevate the Curie temperature. Based on the calculated exchange parameters $J_{ij}$ for both interlayer and intralayer spin configurations, we performed Monte Carlo simulations to estimate the Curie temperature of two representative systems, i.e. bilayer CrI$_3$ on CdSe(0001) and 2D InSe. The specific heat ($C$) was computed based on the Heisenberg model, and the peak on $C(T)$ curve corresponds to the temperature ($T_c$) for second-order magnetic transition [36]. An (30 × 30) supercell was used, and the simulation for each system lasted for $10^5$ loops at each temperature. As shown in Fig. 5(d), $T_c$ is determined to be 130 and 60 K for bilayer CrI$_3$ on CdSe(0001) and 2D InSe, respectively, while $T_c$ for freestanding CrI$_3$ monolayer is 40 K from our calculation (45 K in experiment [9]). Encouragingly, the FM order is strengthened by depositing bilayer CrI$_3$ on bulk semiconductor as $T_c$ increases by up to 3 times.

According to the above discussions, the proximity effect on 2D CrI$_3$ sheet from the semiconducting substrates originates from the interfacial electronic coupling and charge transfer, which greatly reduces the virtual energy gap and modifies the occupancy of $e_g$ orbitals, leading to enhanced $e_g$-$t_{2g}$ hybridization and strengthened interlayer ferromagnetic order in bilayer CrI$_3$. The interfacial charge transfer is correlated to the work function of the substrate relative to that of CrI$_3$. Compared to the widely adopted approaches like heteroatom doping and functionalization with inherent limitations, depositing 2D magnetic materials on proper semiconducting substrates is a universe strategy to manipulate the spin configuration toward robust ferromagnetism with high Curie temperature for practical applications in spintronics.

## 4. Conclusion

In summary, we exploited various types of semiconducting substrates to tailor the interlayer spin configuration of bilayer $CrI_3$. According to our first-principles calculations and Monte Carlo simulations, the antiferromagnetic ground state of $CrI_3$ bilayer in high-temperature phase can be transformed into ferromagnetic state when deposited on suitable bulk surfaces or 2D materials. In particular, the (0001) surface of II-VI semiconductors (CdSe, ZnS and ZnO) form covalent bonds with $CrI_3$ bilayer and endow it robust intralayer and interlayer FM order with Curie temperature up to 130 K, much higher than that on 2D substrates with vdW interaction. Such strong proximity effect can be understood by the prominent charge transfer from II-VI semiconductors, leading to carrier doping to the band structure of $CrI_3$, which significantly enhances the $e_g$-$t_{2g}$ hybridization and strengthens the FM nearest-neighbor exchange. These theoretical results provide essential guidance for obtaining high Curie temperature 2D ferromagnets on semiconducting substrates for practical uses.

**Conflict of interest**

There are no conflicts to declare.

**Acknowledgement**

This work was supported by the National Natural Science Foundation of China (11974068), the Fundamental Research Funds for the Central Universities of China (DUT20LAB110), and the Supercomputing Center of Dalian University of Technology.

TABLE I. Structural and magnetic properties of bilayer CrI$_3$ on various substrates, including lattice mismatch ($\delta$), average interlayer distance between CrI$_3$ and substrates ($d$), interlayer binding energy ($E_{bind}$), work function of substrates ($\Phi$) (experimental values in brackets), charge transfer (CT) from substrates to CrI$_3$, interlayer exchange energy ($E_M$), intralayer exchange energy for the bottom CrI$_3$ layer ($E_{M\parallel}$), and the associated exchange parameters ($J$).

| System | $\delta$ (%) | $d$ (Å) | $E_{bind}$ (eV/f.u.) | $\Phi$ (eV) | CT ($e$/f.u.) | $E_M$ (meV/f.u.) | $J_{1\perp}$ (meV) | $J_{2\perp}$ (meV) | $E_{M\parallel}$ (meV/f.u.) | $J_{1\parallel}$ (meV) | $J_{2\parallel}$ (meV) |
|---|---|---|---|---|---|---|---|---|---|---|---|
| CdSe(0001) | 0.90 | 2.58 | −4.48 | 3.81(3.72 [37]) | 0.24 | 9.42 | 6.32 | −1.76 | 43.31 | 6.42 | 3.69 |
| ZnS(0001) | 0.79 | 2.59 | −2.46 | 3.68(4.10 [38]) | 0.30 | 5.59 | 8.19 | −1.71 | 41.41 | 6.13 | 1.62 |
| ZnO(0001) | 0.12 | 2.13 | −2.32 | 3.60(3.70 [39]) | 0.41 | 5.04 | 6.51 | −2.43 | 38.16 | 5.65 | 1.04 |
| Si(111) | 0.32 | 2.28 | −1.76 | 4.58(4.40 [40]) | 0.19 | 3.26 | 9.19 | −4.00 | 37.29 | 5.52 | 4.15 |
| SnS | 1.16 | 3.29 | −0.41 | 4.33 | 0.06 | 1.08 | 2.84 | −0.61 | 33.96 | 5.03 | 1.28 |
| MoS$_2$ | 0.46 | 3.42 | −0.38 | 5.70 | −0.03 | 0.09 | 1.05 | −0.41 | 22.28 | 3.30 | 1.37 |
| WSe$_2$ | 0.51 | 3.55 | −0.39 | 4.91 | 0.01 | 0.08 | 0.44 | −0.17 | 28.38 | 4.20 | 1.72 |
| GaSe | 0.58 | 3.49 | −0.29 | 5.64 | −0.08 | 0.06 | 5.43 | −2.40 | 18.52 | 2.74 | 1.20 |
| $h$-BN | 0.31 | 3.60 | −0.28 | 5.57 | 0.01 | −0.06 | −1.21 | 2.71 | 19.06 | 2.82 | 1.23 |
| Black P | 1.27 | 3.44 | −0.34 | 4.90(5.04 [41]) | 0 | −0.08 | −10.27 | 4.55 | 28.85 | 4.27 | 2.48 |
| InSe | 1.30 | 3.18 | −0.35 | 5.75 | −0.04 | −0.30 | −19.78 | 0.91 | 17.14 | 2.54 | 1.36 |

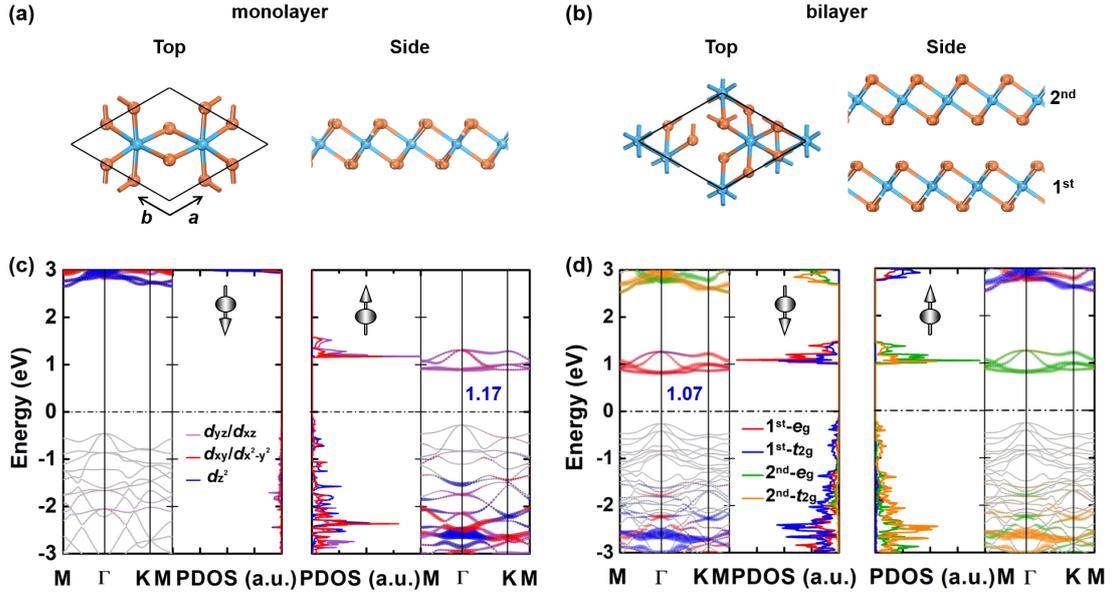

FIG. 1. Atomic structures (top and side views) and spin-polarized electronic band structures along with projected density of states (PDOS) of monolayer (a, c) and bilayer (b, d) $CrI_3$, respectively. The Cr and I atoms are shown in blue and orange colors, respectively. *a* and *b* are the lattice vectors. The colored lines in bands and PDOS are from *d* orbitals of Cr atoms. The blue numbers give the band gap (in eV) for each system. The Fermi level (dashed lines) is set to zero. On each panel of PDOS, the spin orientation is given by an up or down arrow.

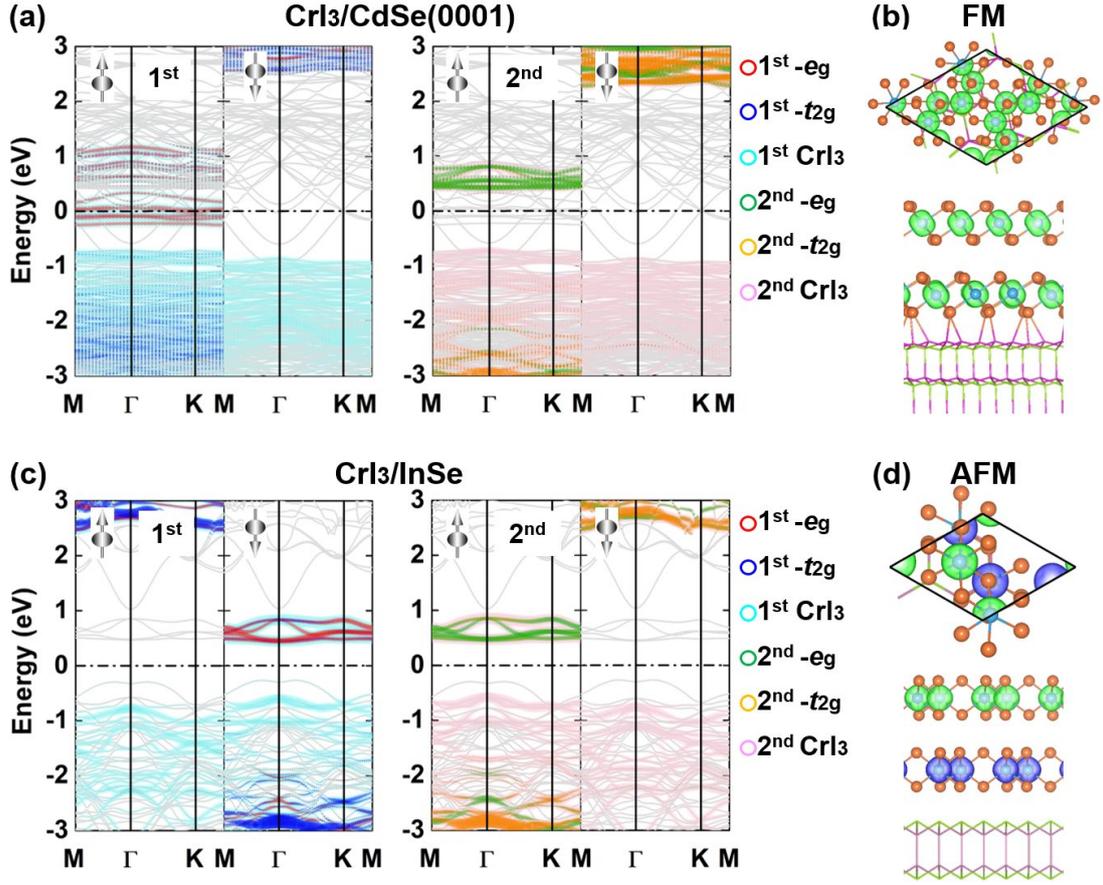

FIG. 2. Spin-polarized band structures of bilayer CrI$_3$ on (a) CdSe(0001) and (c) 2D InSe, respectively. Left panel: 1$^{st}$ CrI$_3$ layer (cyan), right panel: 2$^{nd}$ CrI$_3$ layer (magenta); the bands from substrates are shown in light gray; the projected bands from $e_g$ and $t_{2g}$ orbitals are given in the other colors. The Fermi level (dashed lines) is set to zero. The spin densities of the ground states are shown in (b) and (d). The spin up and spin down components (labeled by arrows) are shown in green and blue colors, respectively, with an isosurface value of 0.01 $e/Å^3$.

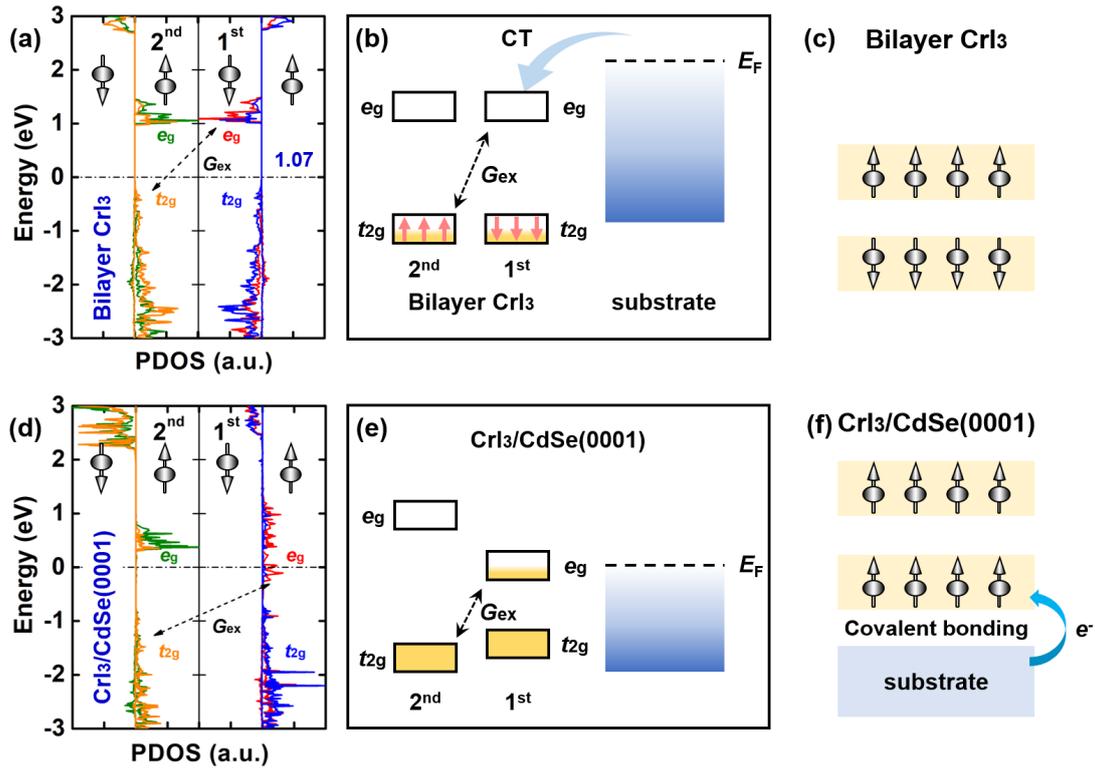

FIG. 3. (a, d) Projected density of states (PDOS) from the $e_g$ and $t_{2g}$ orbitals of freestanding bilayer $CrI_3$ and that on CdSe(0001) surface, respectively. On each panel of PDOS, the spin orientation is given by an up or down arrow. The schematic illustrations of orbital alignments (b, e) and spin configurations (c, f) are shown for each system.

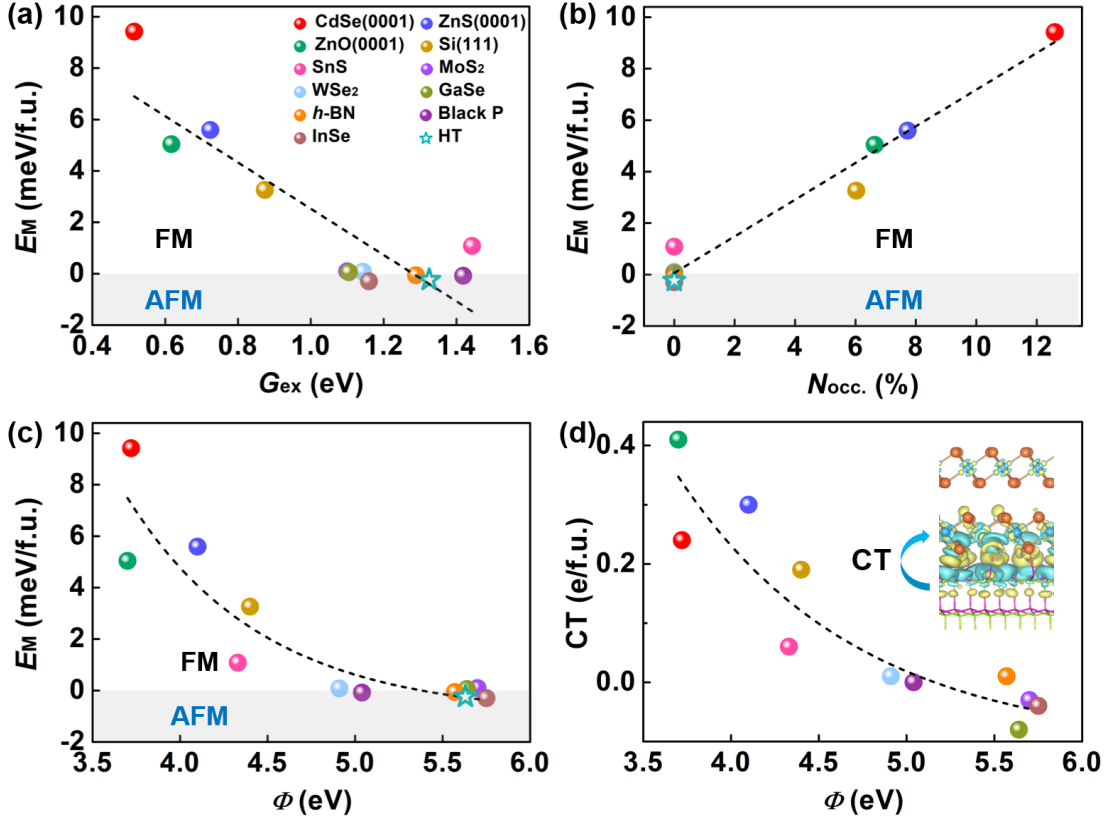

FIG. 4. (a-c) Interlayer FM exchange energy ($E_M$) of bilayer CrI$_3$ on various substrates as a function of virtual exchange gap ($G_{ex}$), occupancy of $e_g$ orbital ($N_{occ.}$), and work function ($\Phi$) of substrates, respectively. The values of freestanding bilayer CrI$_3$ in high temperature phase (denoted as HT) are also given. (d) Interfacial charge transfer (CT) vs. $\Phi$, with the inset showing differential charge densities between bilayer CrI$_3$ and substrate. Yellow and cyan colors represent the charge accumulation and depletion regions, respectively, with an isosurface value of $2\times10^{-4}$ $e$/Å$^3$.

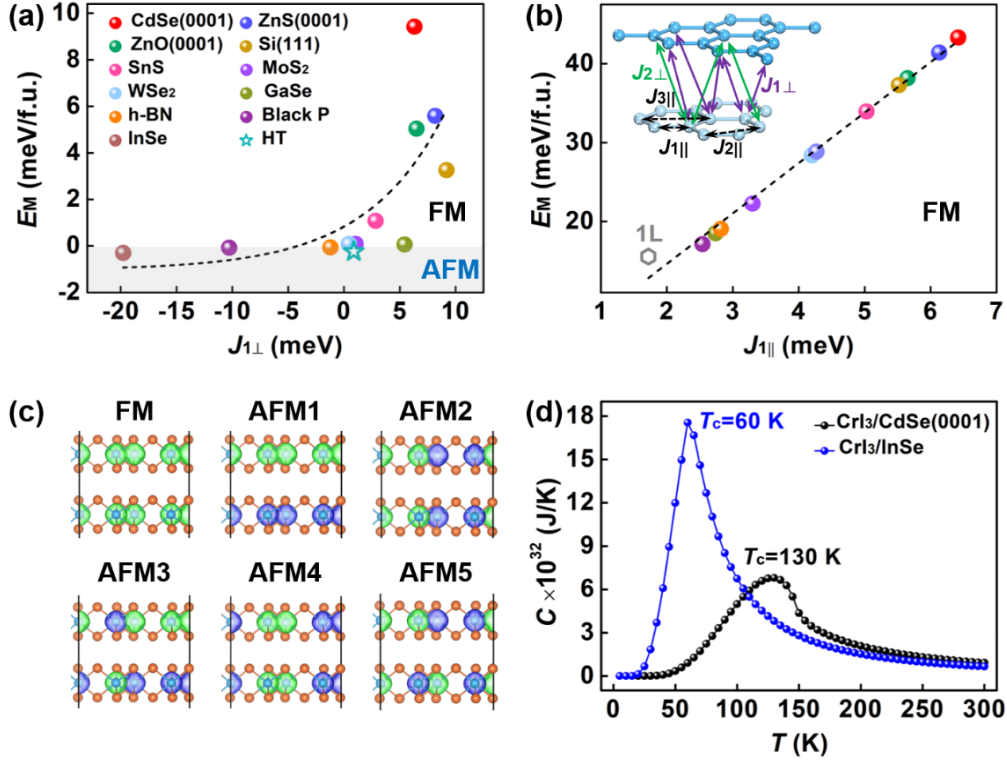

FIG. 5. (a) Interlayer FM exchange energy ($E_M$) of supported bilayer CrI$_3$ as a function of the interlayer nearest-neighbor exchange parameter ($J_{1\perp}$). The values of freestanding bilayer CrI$_3$ in high temperature phase (denoted as HT) are also given. (b) Intralayer FM exchange energy ($E_{M\parallel}$) of the bottom CrI$_3$ layer on substrates as a function of the intralayer nearest-neighbor exchange parameter ($J_{1\parallel}$). The inset displays the five intralayer and interlayer Cr−Cr near neighbors and the corresponding exchange parameters ($J$). (c) Spin densities of the five lowest-energy AFM states in freestanding bilayer CrI$_3$. The spin up and spin down components are shown in green and blue colors, respectively, with an isosurface value of 0.01 $e/Å^3$. (d) Specific heat ($C$) as a function of temperature from Monte Carlo simulation for bilayer CrI$_3$ on CdSe(0001) and 2D InSe, respectively.